\documentclass[11pt]{article}

\usepackage{amssymb,amsmath,amsfonts,amsthm}
\usepackage{latexsym}
\usepackage{graphics}
\usepackage{indentfirst}
\usepackage{hyperref}
\usepackage{comment}
\usepackage[shortlabels]{enumitem}
\usepackage[english]{babel}
\usepackage{tikz}
\usepackage{esdiff}
\usepackage{amsthm}
\usepackage{amssymb}
\usepackage{epsfig}
\usepackage{psfrag}

\newtheorem*{thm*}{Theorem}
\newtheorem{thm}{Theorem}
\newtheorem{lem}[thm]{Lemma}

\newtheorem{cor}[thm]{Corollary}
\newtheorem{fact}[thm]{Fact}

\newtheorem{cl}[thm]{Claim}

\newcommand{\opt}{\mathit{opt}}
\newcommand{\OPT}{\mathit{OPT}}

\allowdisplaybreaks

\makeatletter
\newcommand{\specificthanks}[1]{\@fnsymbol{#1}}
\makeatother

\begin{document}
	
\title{An Improved Algorithm for Finding Maximum Outerplanar Subgraphs}
	
\author{Gruia Calinescu\thanks{Department of Computer Science, Illinois Institute of Technology, Chicago, IL 60616. E-mail: {\tt calinescu@iit.edu }} \and Hemanshu Kaul \thanks{Department of Applied Mathematics, Illinois Institute of Technology, Chicago, IL 60616. E-mail: {\tt kaul@iit.edu}} \and Bahareh Kudarzi\thanks{Department of Applied Mathematics, Illinois Institute of Technology, Chicago, IL 60616. E-mail: {\tt bkudarzi@hawk.iit.edu}} }

\maketitle
	
\begin{abstract}
		
\medskip
We study the NP-complete {\sc Maximum Outerplanar Subgraph} problem. The previous best known approximation ratio for this problem is $2/3$. We propose a new approximation algorithm which improves the ratio to $7/10$.
\medskip



\end{abstract}

\section{Introduction}\label{intro}
A graph is {\em planar}\ if it can be drawn in the 2-dimensional plane such that no two edges meet in a point other than a common end. A planar graph is {\em outerplanar}\ if it can be drawn in a way in which every vertex lies on the boundary of the same connected region of the plane. Let $G$ be a simple graph, and $G'$ be a subgraph of $G$. We say that $G'$ is a {\em maximum outerplanar subgraph}\ of $G$ if it is outerplanar and there is no outerplanar subgraph $G''$ such that $|E(G'')|>|E(G')|$. Given a graph $G$, finding an
outerplanar subgraph of $G$ with the maximum number of edges is called the {\sc Maximum Outerplanar Subgraph} problem.

 While outerplanar graphs have been investigated in-depth for their applications \cite{KANT19961} and theoretical properties \cite{MANNING,felsner,maheshwari,G05,SYSLO197947,KEDLAYA1996238,CGNRS06,OS81}, the problem of finding large outerplanar subgraphs of a graph has not been studied as much.

Most problems which are NP-hard on arbitrary graphs become polynomial on outerplanar graphs\cite{garey,Baker111}. 
An in-depth exploration of practical algorithms for
{\sc Maximum Outerplanar Subgraph} was done by Poranen 
\cite{P05}. 
His main experimental result is that simulated annealing with initial solution taken from the greedy triangular
cactus approximation algorithm (which we discuss below) yields the best known heuristic for the {\sc Maximum Outerplanar Subgraph} problem.

Maximum Outerplanar Subgraph problem is NP-hard, \cite{Yan78,garey}. Hence, instead of solving the problem exactly, we use polynomial-time approximation algorithms.
An algorithm's approximation ratio is the worst-case ratio between the result obtained by the algorithm and the optimal solution. 
 For maximization problems, the approximation ratio is smaller than 1,  and the closer we are to 1, the better.

{\bf Previous Work.} 
By constructing a spanning tree of a given graph (recall, a tree is an outerplanar graph), one obtains an approximation  ratio for the {\sc Maximum Outerplanar Subgraph} problem of 1/2 \cite{cimikowskicoppersmith}.
The approximation ratio was improved by using the concept of {\em triangular cactus}, discussed in Section \ref{prelim} and depicted in Figure \ref{cactus}.
 The greedy triangular cactus approximation algorithm
 from \cite{CalinescuFFK98}
results in a non-trivial approximation ratio of 7/12. 
Furthermore, \cite{CalinescuFFK98} used the fact that
computing a triangular cactus with maximum number of
triangles can be done in polynomial time 
based on {\sc Matroid Parity} \cite{LovPlu,GabSta,S03, CLL14,Orlin08}
to obtain a $2/3$ approximation ratio.

Utilizing an outerplanarity testing algorithm \cite{brehaut77,syslo78,MITCHELL79,syslo79,wiegers}, a greedy technique to approximate the maximum outerplanar subgraph involves adding edges to an outerplanar subgraph as long as the subgraph stays outerplanar.
Cimikowski and Coppersmith \cite{cimikowskicoppersmith} showed that after constructing the spanning tree, using this greedy technique does not improve the approximation ratio of $1/2$.
Poranen \cite{P05} claimed the same for the $7/12$ ratio obtained by the greedy triangular cactus. There are examples that this
holds for the $2/3$ approximation as well.
All these bounds assume that the greedy technique adds the worst possible edge if it has a choice.  In practice, this greedy technique is effective 
\cite{poranen, P05, Poranen_2008}.

{\bf{Related Work.}}
The {\sc Maximum Induced Outerplanar Subgraph} problem is the task of finding the size of the largest subset of vertices in a graph that induces an outerplanar
subgraph. This problem is known to be NP-hard \cite{Yan78} and does not admit good approximations \cite{LY93,FK05}. Morgan and Farr \cite{MF07} 
presented an efficient algorithm that
finds an induced outerplanar subgraph with at least $3n/(d + 5/3)$ vertices
for graphs of $n$ vertices with maximum degree at most $d$.
Donkers et al. \cite{Donkerrrrs}
study the {\sc Outerplanar Deletion} problem, in which one wants to remove at most
$k$ vertices from a graph to make it outerplanar,
and showed that this problem is fixed-parameter tractable.

Another related problem is 
{\sc Maximum Planar Subgraph}, in which one wants to find a
planar subgraph of the input graph with the maximum number of edges.
This problem is also known to be NP-hard \cite{LiuGel}, and it has many applications
(see e.g. \cite{liebers2004,resende,P05,Poranen_2008}).
For this problem, constructing a spanning tree of a given graph
gives an approximation ratio of $1/3$, the greedy triangular cactus approximation algorithm has ratio $7/18$, and the best known approximation is $4/9$ and
is obtained by computing a triangular cactus with maximum number of
triangles \cite{CalinescuFFK98}.

{\sc Maximum Weight Outerplanar Subgraph}, in which one wants to find an
outerplanar subgraph with the maximum weight of the input edge-weighted graph, admits an approximation ratio of $1/2$ by considering the maximum weight spanning tree. This has been improved by \cite{calinescu-new} to 7/12, and by combining Theorem 29 of
\cite{calinescu-new} with the recent algorithm for 
 {\sc Weighted Matroid Parity} \cite{iwata2021weighted}, one 
 immediately obtains a $(2/3)$-approximation algorithm.
 A $(2/3)$-approximation was also claimed in the Master thesis of Osipov \cite{Osipov2006}.

In this paper, we present,
\begin{thm}
\label{t_1}
There exists a polynomial-time $(7/10)$-approximation algorithm for the
{\sc Maximum Outerplanar Subgraph} problem.
\end{thm}
Our algorithm has a greedy phase of adding appropriate induced 4-cycles after the main phase of the
(2/3)-approximation of \cite{CalinescuFFK98}. While the algorithm is very simple (except for the {\sc Matroid Parity} part already used by the
previous best work), the tight analysis we provide is nice and elementary, and may have wider applications.

Applying a greedy method after matching methods ({\sc Matroid Parity} is an
extension of graph matching) is new to us; applying matching methods
after greedy methods was done before in 
\cite{KRY95b}, while \cite{DF97} combines local improvement with matching.

In Section \ref{prelim} we introduce further definitions and notation.
Our algorithm and the proof of Theorem \ref{t_1} appears in Section
\ref{approx-algo}. In Section \ref{s-co-re} we discuss some limitations of our method and conclude with some open questions.

\section{Preliminaries}\label{prelim}

In this paper all graphs are nonempty, finite, simple graphs unless otherwise noted.  Generally speaking we follow West~\cite{W02} for terminology and notation.

Given a graph $G=(V,E)$,  $V'\subseteq V$, and  $E'\subseteq E$, we denote by $G[V']$, the
{\em induced}\ subgraph of $G$ with vertex set given by $V'$, and we denote by $G[E'] = (V,E')$, the
{\em spanning}\ subgraph of $G$ with edge set given by $E'$.  
A {\em triangle}\ in a graph is  $C_3$, a cycle of length three and a {\em square}\ in a graph is an induced $C_4$, a cycle of length four.

A {\em plane graph} is defined as an embedding of an planar graph with a mapping from every vertex to a point in the
2-dimensional plane, and from every edge to a curve on the plane, such that the extreme points of each curve are the points mapped from its end vertices, and all curves are disjoint except on their extreme points. 
The plane graph divides the plane into a set of connected regions, called faces. Each face is bounded by a closed walk called the boundary of the face. The {\em outer face} is the unbounded region outside the whole embedding.
An {\em outerplane graph} is a plane graph where every vertex is mapped to a point on the boundary of the outer face. 
Any other face is called an {\em inner face} of the outerplane graph.

An {\em outeredge}\ in an outerplane graph is an edge which is in the boundary of the outer face and an {\em inneredge}\ in an outerplane graph is an edge which is not an outeredge.
The boundary of the outer face of a biconnected outerplane graph is a cycle.
An {\em inner triangle} is a triangle that is the boundary of an inner face.
An {\em inner square} is a square that is the boundary of an inner face.

Let $H$ be a biconnected outerplane graph and $uv$ be an edge of $H$ and $w\in V(H)\backslash\{u,v\}$. Consider the path $P$ from $u$ to $v$ using only outeredges of $H$ that does not contain $w$ and let $H_1=H[V(P)]$. We say that the subgraph $H_1$ of $H$ is {\em split from $w$ by $uv$}. As an example, in Figure~\ref{case1p1},
$H_1$ is split from $y$ by $xz$. Note that if $uv$ is an outeredge, then $H_1$ will be the graph $\left(\{u,v\},\{uv\} \right)$.

\begin{figure}[th]
\psfrag{(a)}{(a)}
\psfrag{(b)}{(b)}
\begin{center}\leavevmode%
\scalebox{0.57}{
  \includegraphics{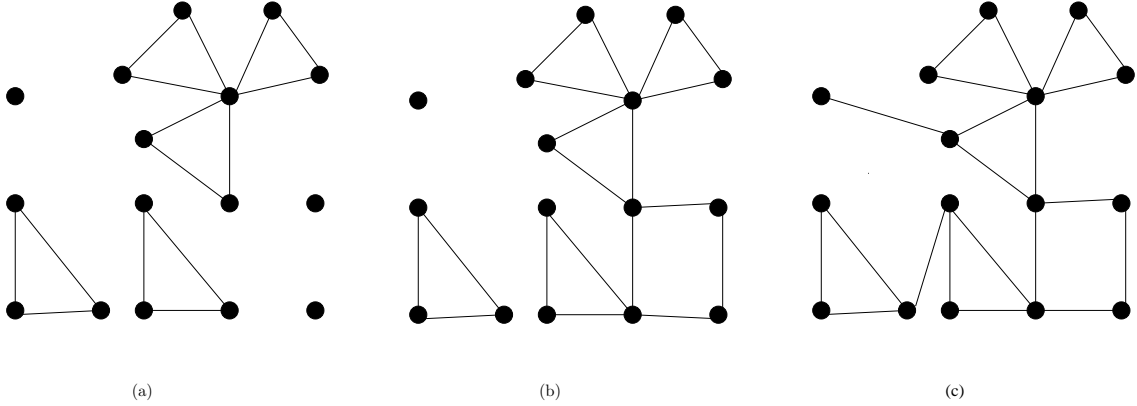}
}
\end{center}
\caption{Figure (a) is a triangular cactus. Figure (b) is a square-triangular cactus. Figure (c) is a square-triangular structure.
}
\label{cactus}
\end{figure}

A {\em triangular cactus}\ is a graph whose cycles (if any) are triangles and
such that all edges appear in some cycle. A {\em triangular cactus in a graph
$G$} is a subgraph of $G$ which is a triangular cactus. A triangular cactus in a graph $G$ is {\em maximum} if it has maximum number of triangles.
A {\em square-triangular cactus}\ is a graph whose cycles (if any) are triangles or squares and
such that all edges appear in some cycle. A {\em square-triangular cactus in a graph
$G$}\ is a subgraph of $G$ which is a square-triangular cactus.
A {\em square-triangular structure}\ is a graph whose cycles (if any) are triangles or squares.  
Note that every square-triangular cactus is a
square-triangular structure, but not vice versa. See Figure~\ref{cactus}.

\section{The Approximation Algorithm}\label{approx-algo}

In this section we describe our algorithm, which we call STS for square-triangular structure, and give its approximation ratio analysis.

Input to the algorithm below is a graph $G$.
\vspace{3mm}

\noindent {\bf Algorithm STS} 

\vspace{3mm}

The algorithm has three phases, as numbered below:

\begin{enumerate}

\vspace{1mm}

\item  Find $M_0$  a {\em maximum}\ triangular cactus in $G$.
\vspace{1mm}

\item Starting with $E_1=E(M_0)$, repeatedly (as long as possible) find 
a square $S$ whose vertices are in different components of $G[E_1]$,
and add the edges of $S$ to $E_1$.
Let $M_1 := G[E_1]$.
\vspace{1mm}

\item Starting with $E_2=E(M_1)$, repeatedly (as long as possible) find an
edge $e$ in $G$ whose endpoints are in different  components of $G[E_2]$, and
add $e$ to $E_2$.

\vspace{1mm}
\noindent Output $G[E_2]$.

\end{enumerate}

Algorithm STS produces a square-triangular structure in the given graph $G$. This is indeed an outerplanar graph.

The $(2/3)$-approximation algorithm of \cite{CalinescuFFK98} has only two phases corresponding to our Phase 1 and Phase 3.

Phase 1 of the algorithm can be implemented to run
in polynomial time as explained in \cite{CalinescuFFK98}.
This is done by an immediate reduction to 
the Graphic Matroid Parity problem, for which 
\cite{LovPlu,GabSta,S03,CLL14,Orlin08} provide polynomial-time algorithms.

Phase 2 can be implemented in time $O(n^4)$ as we try all
possible subsets of four vertices to see if they form a
square that can be added. Within the same time bounds,
we also maintain the connected components
of $G[E_1]$ - these components need to be updated only $O(n)$ times.

To complete the proof of Theorem~\ref{t_1}, it only remains to show the approximation ratio of $7/10$. 

\subsection{Approximation Ratio Analysis}\label{approx}
To establish the approximation ratio, we first prove a lemma about the structure of biconnected outerplanar graphs. This is a generalization of Lemma 3.1 of \cite{CalinescuFFK98}.

\begin{lem}
\label{l_rt}
Let $H$ be a biconnected outerplane graph. Suppose that $H$ has $t$ inner triangles.
Then, the following holds:
\begin{enumerate}
    \item \label{statement1}If $t$ is even, then for every outeredge $xy$ in $H$, there is a triangular cactus $C$ in $H$ with at least $\frac{t}{2}$ triangles such that $x$ and $y$ are in different components of $C$.
    \item \label{statement2}If $t$ is odd, then there is a triangular cactus $C$ in $H$ with at least $\lceil\frac{t}{2}\rceil$ triangles, and for every outeredge $xy$ in $H$, there is a triangular cactus $C$ in $H$ with at least $\lfloor\frac{t}{2}\rfloor$ triangles such that $x$ and $y$ are in different components of $C$.

\end{enumerate}
\end{lem}
\begin{proof}[Proof] 
This proof uses ideas from the proof of
Lemma 3.1 of \cite{CalinescuFFK98}, with many extra cases.
We will prove this by induction on $n+t$, where $n$ is the number of vertices of $H$ and $t$ is the number of inner triangles of $H$.
Note that $n$ is at least 3.
\begin{fact}
\label{fact}
Before we prove the base cases, we note that if a graph $H'$ consists of only one edge $xy$ and therefore it has $t=0$ inner triangles, then there is triangular cactus in $H'$ which has  $r=0 \geq \frac{t}{2}$ triangles and the vertices $x$ and $y$ are in different components of the triangular cactus. The same argument applies when the graph $H$ is biconnected with $n\geq3$ but has zero inner triangles.
\end{fact}

The case  $n+t=3$ cannot happen since if $n=3$ then the biconnected $H$ has
an inner triangle.

Now let $n+t=4$. By Fact~\ref{fact} we can assume that $H$ has one inner triangle $uvw$. Now  $C=H[\{u,v,w\}]$ is a triangular cactus with $r=1$ triangle and we have $r\geq \lceil \frac{t}{2}\rceil =1 $. For the outeredge $xy$, subgraph $C=\emptyset$ is a triangular cactus in $H$ with $r=0$ such that  $x$ and $y$ are not in the same component of $C$. Then we have $r\geq \lfloor\frac{t}{2}\rfloor=0$.
Now assume that the statements are true for any $n'$ and $t'$ such that $n'+t'<n+t$. 
We will prove the statements for $n+t$.

In the first part of the proof, we assume that $t$ is even.

Let $f$ be the inner face such that its boundary contains the edge $xy$. Note that $f$ is not all of $H$ because in that case $H$ must have less than two inner triangles. There are two cases: the boundary of $f$ is a triangle or not.

\emph{Case} 1: Suppose that the boundary of $f$ is a triangle consisting of vertices $x$, $y$, $z$.

    Let $H_1$ be the subgraph of $H$ that is split from vertex $y$ by edge $xz$ and $H_2$ be the subgraph that is split from vertex $x$ by edge $yz$. See Figure~\ref{case1p1}. $H_1\cup H_2$ has $t-1$ inner triangles. Since $t$ is even, WLOG, let $H_1$ have an even number of inner triangles $t_1$ and $H_2$ have an odd number of inner triangles $t_2$. 
\begin{figure}
    \centering
   \begin{tikzpicture}
\draw[color=black, fill=white, thin](-1,0) circle (1.5);
\filldraw[black] (-1,1.5) circle (2pt) node [anchor=south]{$z$};
\filldraw[black] (0,-1.13) circle (2pt) node [anchor=west]{$y$};
\filldraw[black] (-2,-1.13) circle (2pt) node [anchor=east]{$x$};
\path [] (0,-1.13) edge [] node [anchor=west]{$\,\,\,H_2\,\,\,\,\,\,\,\,\,\,$} (-1,1.5);
 \path [] (-1,1.5) edge [] node [anchor=east]{$H_1\,\,\,\,$} (-2,-1.13);
\end{tikzpicture}
   \begin{tikzpicture}
 \draw[color=black, fill=white, thin](-1,0) circle (1.5);
\node[] at (-1,0) {$H_2$};
\node[] at (-1.8,1.8) {$H_1$};
\filldraw[black] (-1,1.5) circle (2pt) node [anchor=south]{$x$};
\filldraw[black] (0,1.13) circle (2pt) node [anchor=west]{$y$};
\filldraw[black] (-2,1.13) circle (2pt) node [anchor=east]{$z$};
\path [] (0,1.13) edge[] node{} (-2,1.13);
\end{tikzpicture}
    \caption{Case 1, $f$ is triangular face $xyz$}
    \label{case1p1}
\end{figure}
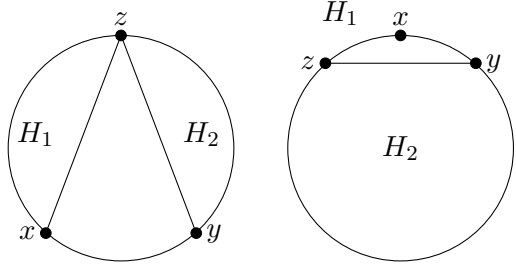
    By applying the induction hypothesis or Fact~\ref{fact} to $H_1$ with outeredge $xz$, there exists a triangular cactus $C_1$ with $r_{1} \geq \frac{t_{1}}{2}$ triangles such that vertices $x$ and $z$ are in different components of $C_1$. By applying the induction hypothesis to $H_2$, there exists a triangular cactus $C_2$ with $r_{2} \geq \lceil\frac{t_{2}}{2}\rceil$ triangles. Since $x$ and $z$ are in different components of $C_1$, we can conclude that $C_1\cup C_2$ is a triangular cactus in which vertices $x$ and $y$ are in different components and has $r$ triangles where  $r=r_1+r_2\geq \frac{t_{1}}{2}+\lceil\frac{t_{2}}{2}\rceil=\frac{t}{2}$.

\begin{figure}[th]
\psfrag{x}{\Huge $x=v_{1_1}$}
\psfrag{y}{\Huge $y=v_{4_2}$}
\psfrag{H'1}{\Huge $H_1$}
\psfrag{H'2}{\Huge $H_2$}
\psfrag{H'3}{\Huge $H_3$}
\psfrag{H'4}{\Huge $H_4$}
\psfrag{e}{$e$}
\psfrag{e1}{\Huge $e_1$}
\psfrag{e2}{\Huge $e_2$}
\psfrag{e3}{\Huge $e_3$}
\psfrag{e4}{\Huge $e_4$}
\psfrag{v12=21}{\Huge $v_{1_2} = v_{2_1}$}
\psfrag{v32}{\Huge $v_{3_2}$}
\psfrag{v22=31}{\Huge $v_{2_2} = v_{3_1}$}
\psfrag{v41}{\Huge $v_{4_1}$}

\begin{center}\leavevmode%
\scalebox{0.40}{
  \includegraphics{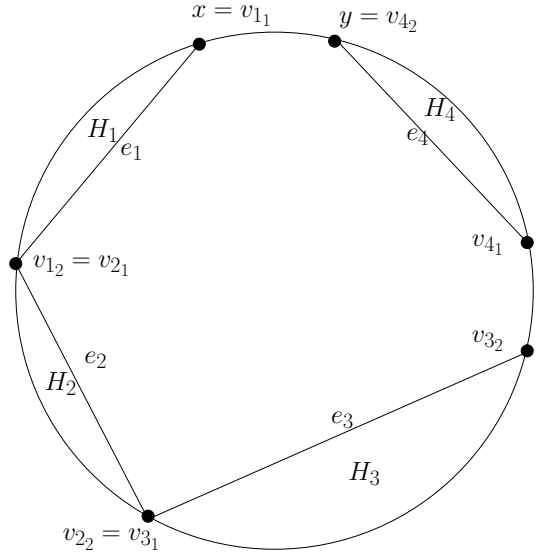}
  }
\end{center}
\caption{
An illustration of Case 2, where straight segments are inneredges of $H$.
}
\label{f_case2}
\end{figure}

   \emph{Case 2}: Suppose that the boundary of $f$ is not a triangle.
    If $t=0$, using Fact~\ref{fact} we are done. From now on in this case we assume that $t>0$.
    Consider all of the inneredges of $f$ and name them as $e_1$, $e_2$, ..., $e_k$. Note that since $t>0$ so $k>0$. For each $i = 1,..,k$, assume that $e_i = v_{i_1} v_{i_2}$ and let vertex $v_i \in V(f)\,\backslash\left \{ v_{i_1}, v_{i_2}  \right \}$. Then, let $H_i$ be the subgraph which is split from vertex $v'_i$ by edge $e_i$.
    See Figure~\ref{f_case2} for an example.
    Assume that $H_i$ has $t_i$ inner triangles.

 For graph $H_i$, we have $t_i+n_i < t+n$ for each $i\in [k]$.

If all $t_i$ are even, by applying induction hypothesis to each $H_i$, we get that there exists a triangular cactus $C_i$ in which $v_{i_1}$ and $v_{i_2}$ are in different components of $C_i$ with  $r_{i} \geq \frac{t_{i}}{2}$ triangles. Now consider the subgraph $C=\bigcup_{i=1}^{k}C_i$ which is a triangular cactus, because the vertices $v_{i_1}$ and $v_{i_2}$ of the edge $e_i$ are in different components of $C_i$ for each $i$, and therefore, in different components of $C$. Hence, $x$ and $y$ cannot be in the same component of $C$.

    If some $t_j$ is odd, then there must be another $t_l$ which is odd. By applying induction on $H_j$, we get that there is a triangular cactus $C_j$ with $r_{j} \geq \lceil\frac{t_{j}}{2}\rceil$ triangles. Then by applying induction on $H_l$ with outeredge (for $H_l$) $e_l$, we get that, there is a triangular cactus $C_l$ which has $r_{l} \geq \lfloor\frac{t_{l}}{2}\rfloor$ triangles such that $v_{l_1}$ and $v_{l_2}$ are not in the same component of $C_l$. For $i \in [k]\setminus \{j,l\}$, by applying induction to $H_i$ there is a triangular cactus $C_i$ with at least $\frac{t_i}{2}$ triangles. Now consider subgraph $C=\bigcup_{i=1}^{k}C_i$ which is a triangular cactus because its cycles can only be triangles and $C$ has $r=\sum_{i=1}^{k}r_i$ triangles and $r\geq \sum_{i\in[k]\backslash \{j,l\}} \frac{t_i}{2}+\lceil\frac{t_{j}}{2}\rceil+\lfloor\frac{t_{l}}{2}\rfloor=\frac{t}{2}$. Note that $x$ and $y$ are not in the same component of $C$ since $v_{l_1}$ and $v_{l_2}$ are not in the same component of $C_l$.
    
    This completes the first part of the proof when $t$ is even.

   For the second part of the proof, we consider that $t$ is odd. We will first prove that there exists a triangular cactus $C$ with $r\geq \lceil\frac{t}{2}\rceil$ triangles.

    Since $t\geq1$ is odd, then $H$ has at least one inner triangle, $uvw$. 
     Let $H_1$ be the subgraph of $H$ that is split from vertex $v$ by edge $uw$, $H_2$ be the subgraph of $H$ that is split from vertex $u$ by edge $vw$, and $H_3$ be the subgraph of $H$ that is split from vertex $w$ by edge $uv$. See Figure~\ref{p2c1}. These three subgraphs together have exactly $t-1$ inner triangles of $H$. 
   Let $H_1$, $H_2$, and $H_3$ have $t_{1}$, $t_{2}$ and $t_{3}$ inner triangles, respectively.  
     Because $t-1$ is even, there are two possibilities: either, each of these three subgraph has even number of inner triangles, or exactly two subgraphs have odd number of inner triangles. 
     
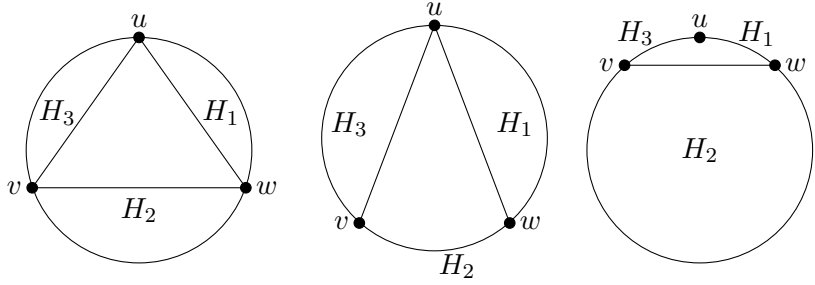
\begin{figure}
    \centering
   \begin{tikzpicture}
\draw[color=black, fill=white, thin](-1,0) circle (1.5);
\node[] at (-0.7,-1.7) {};
\filldraw[black] (-1,1.5) circle (2pt) node [anchor=south]{$u$};
\filldraw[black] (0.42,-0.5) circle (2pt) node [anchor=west]{$w\,\,\,\,\,$};
\filldraw[black] (-2.42,-0.5) circle (2pt) node [anchor=east]{$v$};
\path [] (0.42,-0.5) edge [] node [anchor=north]{$H_2$} (-2.42,-0.5);
\path [] (-1,1.5) edge [] node [anchor=east]{$H_3$} (-2.42,-0.5);
\path [] (-1,1.5) edge [] node [anchor=west]{$H_1$} (0.42,-0.5);
\end{tikzpicture}
   \begin{tikzpicture}
\draw[color=black, fill=white, thin](-1,0) circle (1.5);
\filldraw[black] (-1,1.5) circle (2pt) node [anchor=south]{$u$};
\filldraw[black] (0,-1.13) circle (2pt) node [anchor=west]{$w$};
\filldraw[black] (-2,-1.13) circle (2pt) node [anchor=east]{$v$};
\node[] at (-0.7,-1.7) {$H_2$};
\path [] (0,-1.13) edge [] node [anchor=west]{$\,\,\,H_1\,\,\,\,\,\,\,$} (-1,1.5);
\path [] (-1,1.5) edge [] node [anchor=east]{$H_3\,\,\,\,$} (-2,-1.13);
\end{tikzpicture}
    \begin{tikzpicture}
    \node[] at (-0.7,-1.7) {};
 \draw[color=black, fill=white, thin](-1,0) circle (1.5);
\node[] at (-1,0) {$H_2$};
\node[] at (-0.5,1.55) {$\,\,\,\,\,\,\,\,H_1$};
\node[] at (-1.6,1.56) {$H_3\,\,\,\,\,\,\,\,$};
\filldraw[black] (-1,1.5) circle (2pt) node [anchor=south]{$u$};
\filldraw[black] (0,1.13) circle (2pt) node [anchor=west]{$w$};
\filldraw[black] (-2,1.13) circle (2pt) node [anchor=east]{$v$};
\path [] (0,1.13) edge[] node{} (-2,1.13);
\end{tikzpicture}
    \caption{$t\geq1$ is odd, $H$ has at least one inner triangle, $uvw$.}
    \label{p2c1}
\end{figure}
 For the first possibility, we apply the induction hypothesis or Fact~\ref{fact} to $H_1$ with outeredge $uw$, $H_2$ with outeredge $vw$, and $H_3$ with outeredge $uv$.  For $H_1$, there is a triangular cactus $C_1$ with $r_{1} \geq \frac{t_{1}}{2}$ triangles such that $u$ and $w$ are in different components of $C_1$. Similarly for $H_2$ and $H_3$, there is a triangular cactus $C_2$ with $r_{2} \geq \frac{t_{2}}{2}$ triangles such that $v$ and $w$ are in different components of $C_2$ and there is a triangular cactus $C_3$ with $r_{3} \geq \frac{t_{3}}{2}$ triangles such that $u$ and $v$ are in different components of $C_3$, respectively. We claim that $C=\bigcup_{i=1}^{3}C_i \cup H[\{u,v,w\}]$ is a triangular cactus. Indeed, since $u$, $v$, and $w$ are in different components of each cactus (or they are not in the cacti at all) by putting them together $u,v,w$ are still in different components allowing us to add the triangle $uvw$ and maintain the property of being a triangular cactus. Obviously $C$ has $r=r_1+r_2+r_3+1$ triangles and $t_1
+t_2+t_3=t-1$ which is even, so $r\geq\frac{t_1}{2}+\frac{t_2}{2}+\frac{t_3}{2}+1=\lceil\frac{t}{2}\rceil$.

For the second possibility, WLOG, assume that $t_1$ is even while $t_2$ and $t_3$ are odd. We apply the induction hypothesis or Fact~\ref{fact} to $H_1$ with outeredge $uw$ which tells there is a triangular cactus $C_1$ with $r_{1} \geq \frac{t_{1}}{2}$ triangles such that $u$ and $w$ are in different components of $C_1$. Applying the induction hypothesis 
to $H_2$ and $H_3$, there is a cactus in $H_i$ with $r_{i} \geq \lceil\frac{t_{i}}{2}\rceil$ triangles, for $i=1,2$. Because $u$ and $w$ are in two different components of $C_1$, the subgraph $C=\bigcup_{i=1}^{3}C_i$ is a triangular cactus with $r=r_{1}+r_{2}+r_{3}$ triangles and $r\geq\frac{t_1}{2}+\lceil\frac{t_{2}}{2}\rceil +\lceil\frac{t_{3}}{2}\rceil=\lceil\frac{t}{2}\rceil$.

    Finally, we consider the situation  $t$ is odd and $xy$ be an outeredge. We have proved that there is a triangular cactus $C$ with $r\geq \lceil\frac{t}{2}\rceil$ triangles. If $x$ and $y$ are in different components of $C$ we are done. If $x$ and $y$ are in the same component of $C$ we can remove one of the triangles in that component such that $x$ and $y$ are not in the same component any more. Let $C'$ be the subgraph after removing the triangle. $C'$ is a triangular cactus with at least $\lceil \frac{t}{2}\rceil -1 =\lfloor \frac{t}{2}\rfloor$ triangles.
   This concludes the proof of Lemma \ref{l_rt}. 
\end{proof}

We need the following consequence of the lemma.

\begin{cor}
\label{cor}
Let $H$ be an outerplane graph
with $t$ inner triangles. Then
there exists a triangular cactus in $H$ with $\lceil t/2 \rceil$
triangles.
\end{cor}
\begin{proof}
Assume that $H$ has $k$ components, $H_1, H_2,...,H_k$. Suppose that $H_i$ has $t_i$ inner triangles and $l_i$ blocks, $B_{i1},...,B_{il_i}$.
Using Lemma \ref{l_rt}, every $B_{ij}$ with $t_{ij}$ inner triangles has a triangular cactus $C_{ij}$ as subgraph with at least $\lceil\frac{t_{ij}}{2}\rceil$ triangles. The subgraph $C_i=\bigcup_{j=1}^{l_i} C_{ij}$ is a triangular cactus in $H_i$ since the union does not add any new cycle. $C_i$ has at least $\sum_{j=1}^{l_i} \lceil\frac{t_{ij}}{2}\rceil \geq \lceil \frac{t_i}{2}\rceil $ triangles, since $t_i=\sum_{j=1}^{l_i}t_{ij}$. Now consider subgraph $C=\bigcup_{i=1}^{k} C_{i}$, which is a triangular cactus in $H$ and it has at least $\sum_{i=1}^{k} \lceil\frac{t_{i}}{2}\rceil\geq\lceil t/2 \rceil$ triangles as we have $t=\sum_{i=1}^{k} t_i$.
\end{proof}

 We are now ready to establish the approximation ratio.
\begin{thm} 
\label{t_main}
The approximation ratio of Algorithm STS is $\frac{7}{10}$.
\end{thm}

\begin{proof}
        First we show that the approximation ratio is at most 7/10.
        Take a triangulated outerplanar graph with vertex set $Q$ and $q$ vertices such that $q$ is odd. For each outeredge $uv$ of $Q$, add two new vertices $a_{uv}$ and $b_{uv}$ and the three edges $ua_{uv}$, $a_{uv}b_{uv}$, and $b_{uv}v$ in the outer face of $Q$ such that it remains outerplanar. Name this new graph $H$, it has $3q$ vertices, and $(2q-3) + 3q = 5q -3$ edges. 
Figure~\ref{f_7over10} presents a drawing of $H$ with $q=7$.
Consider $H$ as the input to the algorithm. Note that $H$ is the optimum solution.
 Phase 1 of the algorithm finds a triangular cactus $M_0$  with $\lfloor q/2\rfloor$ triangles connecting all the vertices of $Q$.
 One can easily check that no square can be added to $M_0$ in Phase 2 of the algorithm.
 So the algorithm's output has $3q - 1 + \lfloor q/2\rfloor$ edges,
as every triangle of $M_0$ adds one more edge compared to a spanning tree.
 By letting $q\rightarrow \infty$ the ratio of number of edges in output divided by the number of edges in the optimum converges to $\frac{7}{10}$.

\begin{figure}[th]
\psfrag{(a)}{(a)}
\psfrag{(b)}{(b)}
\psfrag{u}{\Large $u$}
\psfrag{v}{\Large $v$}
\psfrag{auv}{\Large $a_{uv}$}
\psfrag{buv}{\Large $b_{uv}$}
\psfrag{e}{$e$}
\begin{center}\leavevmode%
\scalebox{0.67}{
  \includegraphics{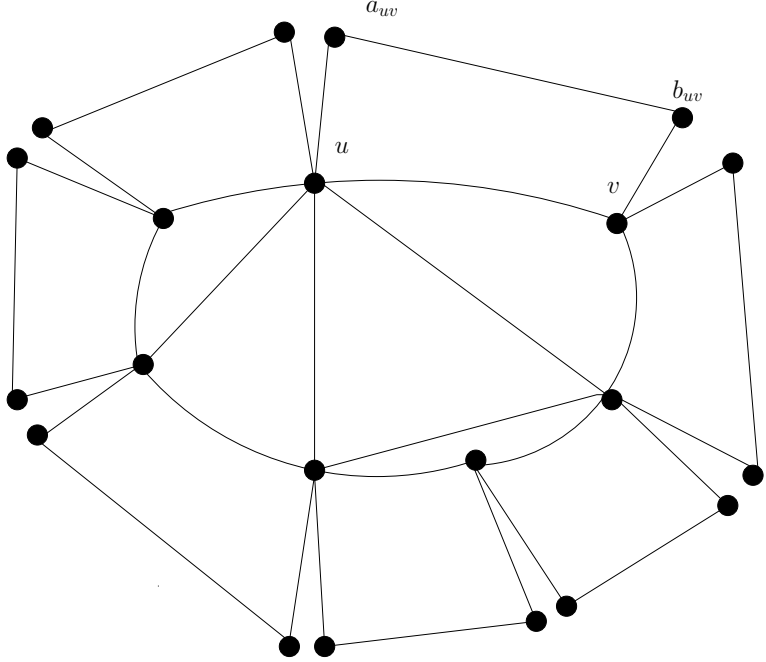}
}
\end{center}
\caption{An example of the graph $H$ used to upper bound the 
approximation ratio of the algorithm.
}
\label{f_7over10}
\end{figure}

Let $\OPT$ denote a maximum outerplanar subgraph of the input graph $G$ and let $\opt$ denote the number of edges of $\OPT$ ($\opt$ is the value of the objective).
We assume that the input graph is connected (or else we can run the algorithm
on every connected component) and therefore $\OPT$ is also connected.
Also, we fix an embedding of $\OPT$, so that it becomes an outerplane graph.
Let $t$ and $s$ denote respectively
the number of inner triangles and inner squares in
$\OPT$.
 Let $r$ and $c$ be respectively the number of triangles and squares 
in the  algorithm's output. 

From Corollary \ref{cor} and the fact that Phase I finds a cactus
with maximum number of triangles in the input graph:
\begin{equation}
t \leq 2r.
\label{eq_cor}
\end{equation}.

We need the following claim.
\begin{cl} 
\begin{equation}
\label{lem2}
    4t+s\leq 10r+6c
\end{equation}
\label{claim_lem2}
\end{cl}
\begin{proof}
Let $M_1$ (from the pseudocode of the algorithm) have $q$ non-trivial connected components $A_1, A_2, \ldots, A_q$.
Let $r_i$, $c_i$ be the number of triangles and squares in $A_i$, respectively.
 This implies that the number of vertices  in the component $A_i$
is exactly $3c_i + 2r_i + 1$.
We have $r = \sum_{j=1}^q r_j$ and $c = \sum_{j=1}^q c_j$.

Let $B_i = \OPT[V(A_i)]$ and let $B_i^j$, $j = 1, 2, \ldots j_i$
 be the connected components of $B_i$, and let $b_i^j  = |V(B_i^j)|$. 
If $b_i^j > 1$, let $B_i^j(l)$, for $l = 1, 2, \ldots l_i^{j}$, be the blocks
of $B_i^j$, and let $b_i^j(l) = |V(B_i^j(l))|$. Note that $b_i^j(l) \geq 2$.
We have
\begin{equation}
\label{vertices}
    \sum_{l=1}^{l_i^{j}} b_i^j(l) = b_i^j + (l_i^{j} - 1).
\end{equation}

Notice that all the inner faces of all the $B_i$ are faces of some
block $B_i^j(l)$.
Consider an inner square $S$ of $\OPT$. At the end of Phase 2 of the algorithm, $S$ will satisfy exactly one of the following conditions.
\begin{enumerate}
\item $S$ is the boundary of an inner face of some $B_i^j(l)$ (we say that $S$ is of Type I)
\item $S$ is not of Type I and
 has one edge that is also an edge of some $B_i^j(l)$ (we say that 
$S$ is of Type II)
\item $S$ is not of Type I or II and 
is such that two non-consecutive vertices
belong to the same $B_i$ (we say that $S$ is of Type III).
\end{enumerate}
If none of these three conditions holds,
 then $S$ can be added to $E_1$ in Phase 2
of the algorithm, and becomes a inner square of Type I. See Figure~\ref{stype}.
Let $s'_1$, $s'_2$, and $s'_3$ be the number of inner squares of $\OPT$ of types I, II,
 and III respectively. Then $s = s'_1 + s'_2 + s'_3$.

No triangle $T$ can be added to $E_1$ while keeping a square-triangular cactus,
since otherwise $M_0$ would not be a triangular cactus with maximum
number of triangles. Using this, we classify the inner triangles of $\OPT$ into
Type I and Type II,
\begin{enumerate}
\item $T$ is the boundary of an inner face of some $B_i^j(l)$ (we say that $T$ is of Type I)
\item $T$ is not of Type I and has at least one edge that is also an
 edge of some $B_i^j(l)$ (we say that $T$ is of Type II).
\end{enumerate}
\begin{figure}[th]
\psfrag{B11(1)}{$B_1^1(1)$}
\psfrag{B11(2)}{$B_1^1(2)$}
\psfrag{B11(3)}{$B_1^1(3)$}
\psfrag{B12(1)}{$B_1^2(1)$}
\psfrag{B12(2)}{$B_1^2(2)$}
\psfrag{T1}{Type I}
\psfrag{T2}{Type II}
\psfrag{T3}{Type III}
\psfrag{T'1}{Type I}
\psfrag{T'2}{Type II}
\psfrag{buv}{\Large $b_{uv}$}
\psfrag{e}{\Large $e$}
\begin{center}\leavevmode%
\scalebox{0.67}{
  \includegraphics{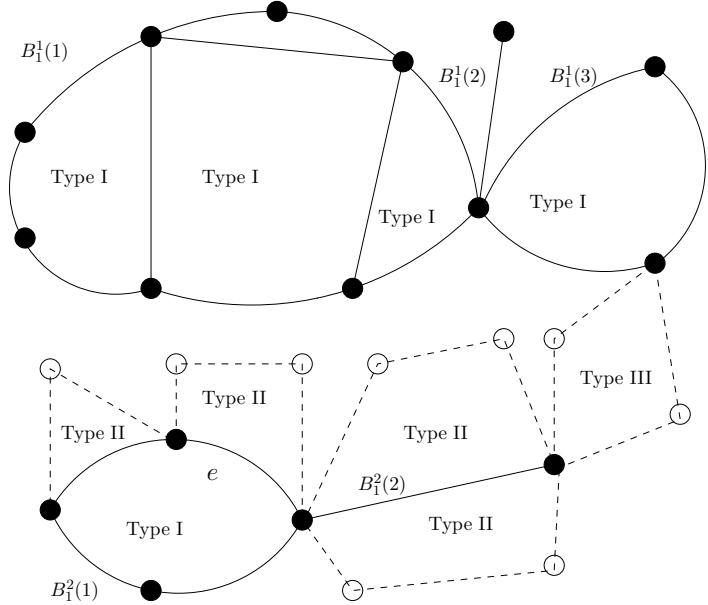}
}
\end{center}
\caption{An example of the different type of inner squares. Filled black circles are vertices in $B_1$ and empty circles are vertices not in $B_1$.
Here we have two connected components of $B_1$: $B_1^1$  and $B_1^2$.
Furthermore, $B_1^1$ has three blocks: $B_1^1(1)$ with 8 vertices, $B_1^1(2)$ with only two vertices, and $B_1^1(3)$, with three vertices. Similarly, $B_1^2$ has two blocks:$B_1^2(1)$ and $B_1^2(2)$.
}
\label{stype}
\end{figure}
Let $t'_1$ and $t'_2$ be the number of inner triangles of $\OPT$ of types I
 and II respectively. Then $t = t'_1 + t'_2$.

Block $B_i^j(l)$ has at most $b_i^j(l)-2$ inner faces. We obtain:
\begin{equation}
\label{e_inner}
t'_1 + s'_1 \leq \sum_{i,j,l} (b_i^j(l)-2).
\end{equation}

We associate with each inner triangle (inner square, respectively) of $\OPT$ of Type II
an edge of some $B_i^j(l)$, precisely the edge that is both
in $B_i^j(l)$ and in the inner triangle (inner square, respectively).

If $b_i^j(l) > 2$, there are exactly $b_i^j(l)$ edges on the
outer face  of $B_i^j(l)$. Any such edge $e$
can be associated with at most one
inner triangle or inner square of Type II, as we argue below. For illustration look at edge $e$ in Figure~\ref{stype}.
Every edge of $\OPT$
participates in at most two inner faces of $\OPT$. 
On one side of $e$
lies an inner face of $B_i^j(l)$ and this is 
also an inner face of $\OPT$, based on the following fact.

\begin{fact} 
Let $H$ be a
outerplane graph. Let $H'$ be an induced subgraph of H. Then, an inner face of $H'$ is also an inner face of $H$. 
\end{fact}

Thus one of the at most two inner triangles or inner squares
of $\OPT$ that
could be associated with $e$ is in fact an inner triangle 
or an inner square of Type I, and not of Type II.

Inneredges of $B_i^j(l)$ are not associated with
any inner triangles or inner squares of Type II, since the boundary of two faces
of $\OPT$ on 
both sides of such an inneredge  of $B_i^j(l)$ 
are of Type I, based on the reasoning above.

If $b_i^j(l) = 2$, there exists exactly one edge in $B_i^j(l)$ and this edge
can be associated with at most two inner triangles or inner squares 
of Type II,
since every edge of $\OPT$
participates in at most two inner faces of $\OPT$. As an illustration, please look at $B_1^2(2)$ in Figure~\ref{stype}.

Based on these arguments, we conclude:

\begin{equation}
\label{e_outer}
t'_2 + s'_2 \leq \sum_{i,j,l} b_i^j(l).
\end{equation}

We continue by counting the inner squares of Type III of $\OPT$.
Let $S$ be such a square with vertices  $u,v,w,x$ in clockwise order.
We must have that some $B_i$ contains two non-consecutive
of these four vertices, and does not contain the other two,
or else $S$ would be of Type I or Type II.
WLOG, let $u,w \in V(B_i)$ and $x,v \not \in V(B_i)$. Note that $u$ and $w$
are not adjacent in $\OPT$, since there is no edge inside this square, and an
edge outside this square will make
 either $x$ or $v$ not on the
outer face of $\OPT$, contradicting outerplanarity.
By the same reasoning, 
$\OPT$ does not have a path from $u$ to $w$ that does
not pass through either $x$ or $v$.
So $u$ and $w$ are in different components of $B_i$.
We associate with $S$ the pair of vertices $u,w$ (recall that $uw$ is not an edge in $\OPT$).
We claim that the total number of inner squares of Type III of $\OPT$
whose corresponding pair of vertices are in different components of $B_i$ is at most
$j_i - 1$ (recall that $j_i$ is the number of components of $B_i$), as we argue in the next paragraph.

Let $F_i$ denote the set of all edges $uw$ for each pair of vertices $u,w$ of $B_i$ corresponding to some square of Type III. $\OPT$ with $F_i$ added is still
outerplanar, and cannot contain a path from $u$ to $w$ (except for the
edge $uw$) that does not go through the corresponding $v$ or $x$. Since
$v$ and $x$ are not added to $B_i$, if we add $F_i$ to $B_i$, then this new graph $(V(B_i),E(B_i)\cup F_i)$ does not contain any cycle which is not in $B_i$. We conclude that $|F_i|\leq j_i-1$. 

Thus, it follows that

\begin{equation}
\label{e_cross}
s'_3 \leq \sum_{i=1}^q (j_i - 1).
\end{equation}
By adding up Equation (\ref{e_outer}) and Equation (\ref{e_inner}) we get that
\begin{equation}
\label{ineq6}
t'_1+s'_1+t'_2+s'_2\leq \sum_{i=1}^{q} \sum_{j\in [j_i],b_i^j>1}\sum_{l=1}^{l_i^j} (2b_i^j(l)-2).
\end{equation}
Using Equation (\ref{vertices}), we have that 
$\sum_{l=1}^{l_i^j} (2b_i^j(l)-2)=2b_i^j-2$. Therefore, Inequality (\ref{ineq6}) simplifies to
\begin{equation*}
   t'_1+s'_1+t'_2+s'_2\leq \sum_{i=1}^{q} \sum_{j\in [j_i],b_i^j>1} (2b_i^j-2),
\end{equation*}
Since for $b_i^j=1$ we have $2b_i^j-2=0$ we can add them to the sum without changing the amount. Hence,
\begin{equation*}
   t'_1+s'_1+t'_2+s'_2\leq \sum_{i=1}^{q} \sum_{j\in [j_i]} (2b_i^j-2).
\end{equation*}
Note that since $B_i^j$ are components of $B_i$ and the set of vertices of $B_i$ is the same as the set of vertices of $A_i$, we will have  $\sum_{j\in [j_i]} b_i^j= |V(B_i)|=|V(A_i)|$ which, recall, is equal to $3c_i+2r_i+1$. Hence,
\begin{equation*}
   t'_1+s'_1+t'_2+s'_2\leq \sum_{i=1}^{q}( 2( 3c_i+2r_i+1)-2 j_i).
\end{equation*}
Now by adding Equation (\ref{e_cross}) to equation above we get that
\begin{equation*}
   t'_1+s'_1+t'_2+s'_2+s'_3\leq \sum_{i=1}^{q}( 2( 3c_i+2r_i+1)-2 j_i + j_i -1).
\end{equation*}
Recall that $t=t'_1+t'_2$ and $s=s'_1+s'_2+s'_3$, hence,
$$t+s\leq \sum_{i}( 2(3c_i+2r_i)-j_i+1)\leq \sum_i (4r_i+6c_i)=4r+6c.$$
Multiplying Equation (\ref{eq_cor})
by 3 and adding it to equation above, we
obtain Equation (\ref{lem2}):
\begin{equation*}
    4t+s\leq 10r+6c.
\end{equation*}
This concludes the proof of Claim \ref{claim_lem2}
\end{proof}

We continue with the proof of Theorem \ref{t_main}. 

Let $\OPT$ have blocks $H_1, H_2, \ldots, H_k$, and let $n_j = |V(H_j)|$
and $m_j = |E(H_j)|$. 
Recall that $\OPT$ is connected.
Then $n := |V| = |V(G)| = |V(\OPT)|
 = \left( \sum_{j=1}^k n_j   \right) - (k-1)$.
We also have that $\opt = |E(\OPT)| = \sum_{j=1}^k m_j$.
Let $f_i^j$ be the number of internal faces of length $i$ in $H_j$,
and $t_j$ and $s_j$ be respectively $f_3^j$  and $f_4^j$.
We have that $n_j = 2 +  t_j + 2 s_j + \sum_{i \geq 5} (i-2) f_i^j$,
and $m_j =  1 +  2 t_j + 3 s_j + \sum_{i \geq 5} (i-1) f_i^j$.

Note that Phase~3 makes the output subgraph connected. The algorithm outputs a solution with $(n-1) + r + c$ edges, as
every triangle or square increases the cyclomatic number\footnote{the cyclomatic number (also called circuit rank, cycle rank, or nullity) of an undirected graph  is the minimum number of edges that must be removed from the graph to break all its cycles, making it into a tree or forest.} of the
output by $1$. 
We have
\begin{flalign*}
n-1+r+c & = r + c  -k + \sum_{j=1}^{k}n_j \\
 & =
r + c  -k  + \sum_{j=1}^{k} \left( 2 +t_j+2s_j+ \sum_{i \geq 5} (i-2) f_i^j \right) \\
 & = 
 r+c + k+t+2s+ \sum_{j=1}^k \sum_{i \geq 5} (i-2) f_i^j ,
\end{flalign*}
and $$\opt=\sum_{j=1}^{k} \left(1+2t_j+3s_j+\sum_{i \geq 5} (i-1) f_i^j \right) = k+2t+3s+\sum_{j=1}^k 
\sum_{i \geq 5} (i-1) f_i^j.$$
For $i\geq 5$, we get that $10(i-2)\geq7(i-1)$.
By Inequality (\ref{lem2}), we have $10r+10c\geq4t+s$ and we also have $10 k\geq 7k$. Hence, one can easily check that
\begin{equation*}
    10 \left( r +c + k+t+2s + \sum_{j=1}^k \sum_{i \geq 5} (i-2) f_i^j \right) 
    \geq  7 \left(k+2t+3s+ \sum_{j=1}^k \sum_{i \geq 5} (i-1) f_i^j \right).
\end{equation*}
Therefore we showed that
$\frac{n-1+r+c}{\opt}\geq \frac{7}{10}$. This finishes the proof of Theorem \ref{t_main}.\
    \end{proof}
    
\section{Concluding Remarks}\label{s-co-re}
In this paper we presented an algorithm which improved the ratio of {\sc Maximum Outerplanar Subgraph} to $7/10$. It is natural to ask if this algorithm can be improved. Some obvious modifications to the algorithm do not help.

Observe that adding pentagons or larger outerplanar graphs after Phase 2 of Algorithm STS will not make the ratio better than $7/10$. The example used in the first part of the proof of Theorem~\ref{t_main} illustrates this. See Figure~\ref{f_7over10}. There are no pentagons or any other structure in the optimum that have their vertices in different components of the graph $M_1$ from the algorithm.

We believe that the greedy technique of adding edges to an outerplanar subgraph as long as the subgraph stays outerplanar does not improve the ratio provided that the worst possible edge is chosen.

An outerplanar $k$-restricted structure is a simple graph whose blocks are outerplanar and each has at most $k$ vertices. In this paper, outerplanar $4$-restricted structures are used by our approximation algorithm. The outerplanar $k$-restricted ratio is the infimum, over simple outerplanar graphs H, of the ratio of the number of edges in a maximum $k$-restricted structure subgraph of $H$ to the number edges of $H$. It is proved in \cite{CF08} that, as $k$ tends to infinity, the outerplanar $k$-restricted ratio tends to $1$. 
This could be useful in improving the
approximation ratio for 
{\sc Maximum Outerplanar Subgraph},
although  we do not know if there is a polynomial time algorithm for 
finding an outerplanar 4-restricted structure with maximum number of edges that is a subgraph of the input graph.

A {\emph diamond} is a cycle of size four with exactly one chord and is an outerplanar graph. Note that a diamond is not a square-triangular structure. 
A diamond has five edges and four vertices, therefore, it is a better structure than a square since it has more edges and also a better structure than two triangles since it has fewer vertices. Since $K_4$ is not outerplanar, the only blocks that an outerplanar $4$-restricted structure can have are  bridges, triangles, squares and diamonds.

\begin{figure}[t]
\begin{center}\leavevmode%
\scalebox{0.30}{
  \includegraphics{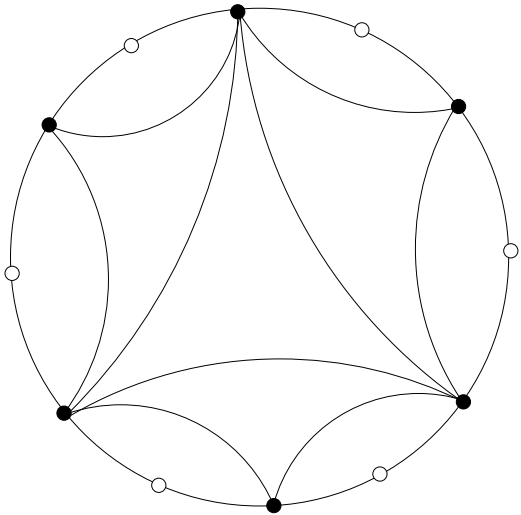}
}
\end{center}
\caption{The vertices of $H_1$ are solid, and the newly
added vertices of $H_2$ are the empty small circles.}
\label{f_gd}
\end{figure}
One way to use diamonds is to add them greedily (as long as the four vertices are in different connected components), followed by triangles as in \cite{Schmid17}. We could also add squares and pentagons after this. And after this,
one employs Phase 3 of Algorithm STS.
This approach does not lead to an approximation ratio better than $2/3$, as the following series of examples shows.
The optimum solutions $H_k$ are
taken from Theorem 4 of \cite{CF08}.
and $H_2$ is illustrated in Figure~\ref{f_gd}. 
$H_0$ is a triangle.
To obtain $H_k$ from $H_{k-1}$ duplicate and 
then subdivide every outeredge.  
Thus $H_k$ has $3 \cdot 2^k$ vertices and is triangulated with
$2 \cdot 3 \cdot 2^k - 3$ edges. We also have a separate 
diamond cactus $D_k$
with $d$ diamonds that includes all the vertices
of $H_k$ that came from $H_{k-1}$, plus another one vertex.
Here, $d = 2^{k-1}$, as one diamond structure with $d$ diamonds has $1 + 3d$ vertices. 
Given an input that has both the edges of $H_k$ and of $D_k$,
the algorithm can select $D_k$ by greedily adding diamonds,
after which it goes directly to Phase 3 as the vertices of $H_k$ that are not
in $D_k$ form an independent set and therefore cannot make
any triangle, square, pentagon, or larger block that
one can greedily add to $D_k$. The output of this greedy algorithm has $3 \cdot 2^k - 1 + 2d$ edges, as every
diamond increases the cyclomatic number by 2.
Recall that $d = 2^{k-1}$ and that the optimum has $2 \cdot 3 \cdot 2^k - 3$ edges. The ratio of the output to optimum converges to $2/3$ as $k$ goes to infinity.
 
The question of how large
an approximation ratio for the {\sc Maximum Outerplanar Subgraph} problem can be achieved remains open.
Is there a linear-time  algorithm
with approximation ratio $1/2+\epsilon$? 
Is there an approximation algorithm for the {\sc Maximum Weight Outerplanar Subgraph} problem with a ratio better than $2/3$?

\end{document}